**A prelude to the multibeam photoacoustics**

Pawel Rochowski

*Institute of Experimental Physics, Faculty of Mathematics, Physics and Informatics, University of Gdańsk, Wita Stwosza 57, 80-308 Gdańsk, Poland*

Correspondence: pawel.rochowski@ug.edu.pl

## Abstract

The research focuses on the introduction and validation of the lock-in photoacoustics of condensed systems in the presence of two simultaneous probing beams. The experiments, performed for three model systems: a perfect absorber - carbon black membrane, a multilayer system – acrylic paint-covered copper plate, and luminescent sample – ruby, were aimed at the comparative analyses of frequency domain photoacoustic responses to a single probe beam (classical approach) and two probe beams (of different wavelength). A modulation frequency shift criterion, allowing for the simultaneous acquisition of two independent signals, was recognized. No synergistic or cross effects of the dual beam excitation on the photoacoustic signal generation (impacts of the beams intensities or modulation frequency effects) were not observed. The multibeam photoacoustic data appear to be (at least) equivalent to classically obtained results for stationary systems. It is considered, that the multibeam approach will extend the applicability of photothermal methods to the time-dependent systems, where the evolution of distinct absorption bands is of special interest. An application of the developed method, related to the site-dependent energy storage efficiency of photosynthetically-active specimen, was described and validated upon three-beams experiment performed on a model *Ficus benjamina L.* leaf.

## 1. Introduction

Due to a specific relation between the thermal and optical properties of materials and the measureable sample response to a modulated probe light, photoacoustics (together with other photothermal techniques, like photothermal radiometry or photothermal spectroscopy) has found numerous applications in applied, biomedical and environmental sciences [1]. Typically, photoacoustic (PA) measurements are performed in the wavelength domain (varying wavelength $\lambda$, constant modulation frequency $f$; hereafter: PhotoAcoustic Spectroscopy - PAS), or in the modulation frequency domain (constant λ, varying *f*; hereafter PhotoAcoustic Depth-Profiling: PADP); depending on the sample, the two modalities can provide (under a certain theoretical frameworks) a direct access to a wide range of characteristics. These include: spectral properties (generally, the PA response proportional to the product of the sample non-radiative deactivation efficiency and light absorption coefficient), radiative quantum efficiency, thermal diffusivity/conductivity (or estimation of layer thicknesses in layered samples of known diffusivity), non-radiative relaxation/carrier recombination times, (time-dependent) pigment distribution within a sample, efficiency of photosynthesis and many more [2–4]. In some cases, PA studies involve two irradiation sources; for example: 1) the determination of photosynthesis efficiency is based on two separate measurements, that is PADP in the presence and in the absence of non-modulated, photosynthesis-saturating white light [5]; 2) the determination of quantum efficiency of luminescent samples based on PA phase analyses, with two separate PADP experiments with distinct *pumping* beams, allowing the effects of saturation and heat flow to be eliminated, as shown by Quimby and Yen [6]; 3) measurement of the energy-resolved distribution of electron traps involves absorption studies using a monochromatic single modulated beam in the presence of wavelength-scanned continuous, non-modulated light (method referred to as Reversed Dual-Beam PAS) [7]. The above examples of two-beams (or dual-beam) PA involve situations, where only one beam (at a time) can actually be considered as the probing beam; even in the case of two simultaneous light sources, a continuous source provides a certain, global change to a system (i.e. saturates specific processes), while the modulated source probes the changed system.

In the broader context of photothermal sciences, heterodyne sensing technique can also be mentioned. In this modality, two simultaneous probing beams of the same wavelength but with different modulation frequencies periodically perturb a certain feature of a system (e.g. produce thermal or carrier density waves), the signature of which is transferred to a detector. In a typical configuration, that is where the sinusoidal probing beams are of similar intensity, one expects the signal to be given by a mix of fundamental and beat frequencies components. Eventually, the heterodyne sensing aims for the signal detection at the beat frequencies, which can be tailored to the capabilities of the detector. The concept has been used, among others, in lock-in carrierography, but also in photoacoustic detection of gases and dual-comb PA spectroscopy [8–10].

According to the author’s survey, there are no reports on a PADP modality with two (or more) probe beams, allowing for the simultaneous analysis of two spectral sites for the same sample.

The research focuses on an introductory and quantitative experimental tests of *f*-domain lock-in photoacoustics in the presence of two simultaneous probing beams illuminating a sample. Attention is paid to the reproducibility and correspondences between two signals (responses to excitations at different wavelengths) recorded separately (one beam at a time), and recorded simultaneously (two beams together). To avoid any confusion with cases already studied in the literature, the method explored here is referred to as the MultiBeam PhotoAcoustics – MBPA.

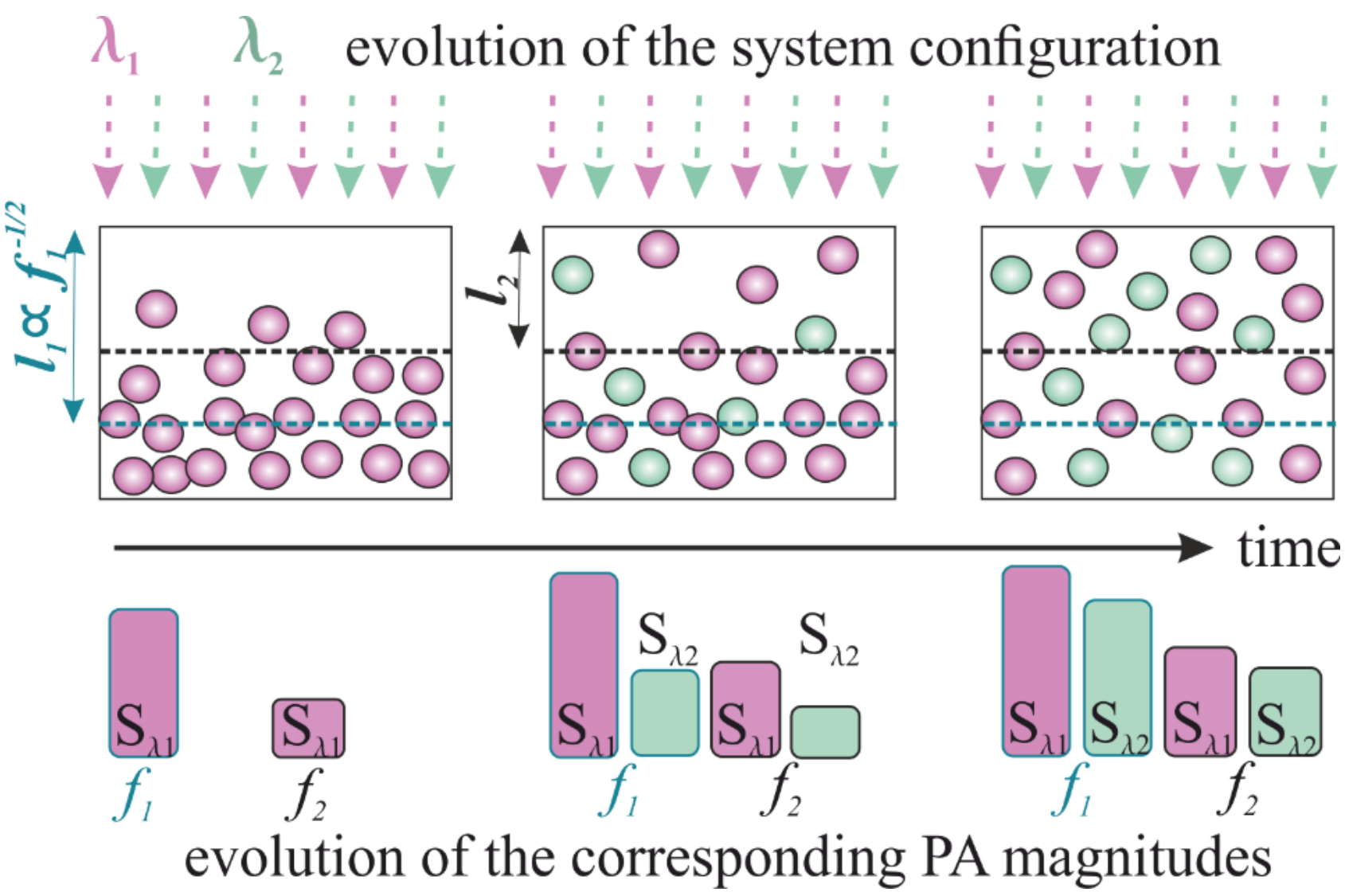

Fig. 1 An exemplary transient system for MBPA studies, consisting of non-uniformly distributed and photosensitive *magenta* particles undergoing time-dependent processes: diffusion and photodegradation (i.e. mass transport and decay problems). The PA setup includes two scanning beams of different wavelengths (corresponding to the examination of *magenta* particles and the photoproduct - *green* particles) operating at two modulation frequencies limiting the sampling depth. The output signals are given by $S(\lambda_{1,2}, f_{1,2})$. By relating each signal of modulation frequency $f_{1,2}$ to the specific sampling depth $l_{1,2}$ ($S(f_{1,2}) \equiv S(l_{1,2})$, black and blue dotted line), it is possible to calculate the local signal contribution $S_L$ for each specimen simply by subtracting the signals from two neighbouring layers, $S_{L_{12}} = S(l_1) - S(l_2)$, where $S_{L_{12}}$ is proportional to the pigment content in the $l_1 - l_2$ layer [2]. It is expected, that the analyses of the PA signal magnitudes can provide information on the mass transport and decay time constants.

The introduction of the MBPA is motivated by the recent need to extend the PA-based methodologies towards the study of transient systems. To highlight the issue, let us consider a condensed sample with an addition of initially non-uniformly distributed pigment molecules undergoing diffusion (i.e. magenta particles in a box, as shown in Fig. 1). One way to describe the evolution of the pigment distribution in the system is to continuously monitor the pigment content (the system response is assumed to be proportional to the number of magenta particles) within a given layer (or layers) of the sample.

In some cases this can be achieved by a proper PA experimental setup, i.e. where the beam wavelength $\lambda$ ($\equiv \lambda_1$) matches the spectral characteristics of the pigment, and the light modulation frequency $f$ is adapted to a certain sampling depth $l$ of interest (the sample layer contributing to the PA response is assumed to follow the relation $l \propto f^{-1/2}$, and depends on the sample thermal diffusivity; as an example, in the frequency range of 10 Hz – 10 kHz, exemplary sampling depths for water-based systems cover the range of ~ 2 – 68 μm. More information on the origin of the relation is provided at the end of section 2: Photoacoustics with solids).

The experiment can easily be extended into pigment content analyses in subsequent sample layers, if an appropriate set of modulation frequencies $f_n$ is used. The situation becomes complicated if the pigment is chemically unstable (i.e. temperature effects, photodegradation etc.), and the product and reactant have different spectral characteristics (magenta particles transform into green particles, as in Fig. 1). Then, the PA response to a single beam excitation is no longer proportional to the total number of migrating particles. The problem can be solved by applying a second scanning beam of wavelength $\lambda_2$. However, such a scenario requires that the PA responses are acquired independently of each other and are proportional to the number of photoacoustics-active particles. This raises to potential issues related to the proper selection of the probe beams modulation frequencies and the presence of secondary effects affecting the individual responses, which need to be investigated.

At first sight, the example of the transient system shown above may seem exotic. On the other hand, it corresponds to the mass transport problems already investigated by photoacoustic methods in applied sciences, i.e. transdermal drug delivery in pharmaceutical sciences [11–13]. The recent results obtained by the author in the mass transport studies further highlighted the need for the MBPA development, since a method that allows the detection of a pigment and its derivatives (more precisely, the evolutions of the pigments distributions inside a membrane) is a must for a thorough and precise experimental study of the principles governing the mass transport and interfacial mass transfers [14]. Moreover, the multibeam PA measurement strategy may be beneficial for the study of other reactive or decaying multispecies systems (a discussion on the relation between the measurement time constant vs dynamical process time constant in PA experiments can be found in [2]). For the purpose of this research, another example of the applicability of MBPA will be studied: the energy storage investigation of an photosynthetically-active system.

The paper is organised as follows: first, an outline of the photoacoustic signal generation in the presence of single beam excitation (the Rosencwaig-Gersho (RG) theory) is provided. Second, a set of PA experimental data under selected conditions (beams intensities/modulation frequencies) on absorbing systems in the presence of single and double beam excitation is presented, including raw and processed data analyses. Third, a discussion of MBPA signal generation in relation to the RG theory is provided. Finally, an example of the MBPA implementation is presented.

## 2. Photoacoustics with solids – basics of the Rosencwaig-Gersho theory

The principles of the photoacoustic signal generation (single excitation beam mode) in solids, are provided by the Rosencwaig and Gersho *piston* (RG) model, developed about 50 years ago [15]. The authors considered a 1D problem of a sample ($x \in< -l_s; 0 >$, where the subscript *s* denotes the sample) set on a backing material ($x \in< -(l_s + l_b); -l_s)$, *b* is for the backing) in a closed cell filled with gas (the gas column occupies $x \in (0; l_g >$, where *g* is for the gas). The PA signal generation was assumed to consists of three subsequent steps: (*i*) absorption of modulated energy flux followed by thermal deactivation of a sample, (*ii*) evolution of the perturbed temperature field in the sample, (*iii*) acoustic signal generation due to periodic heat flow from solid to gas. A brief summary of the RG approach is given in figure 2.

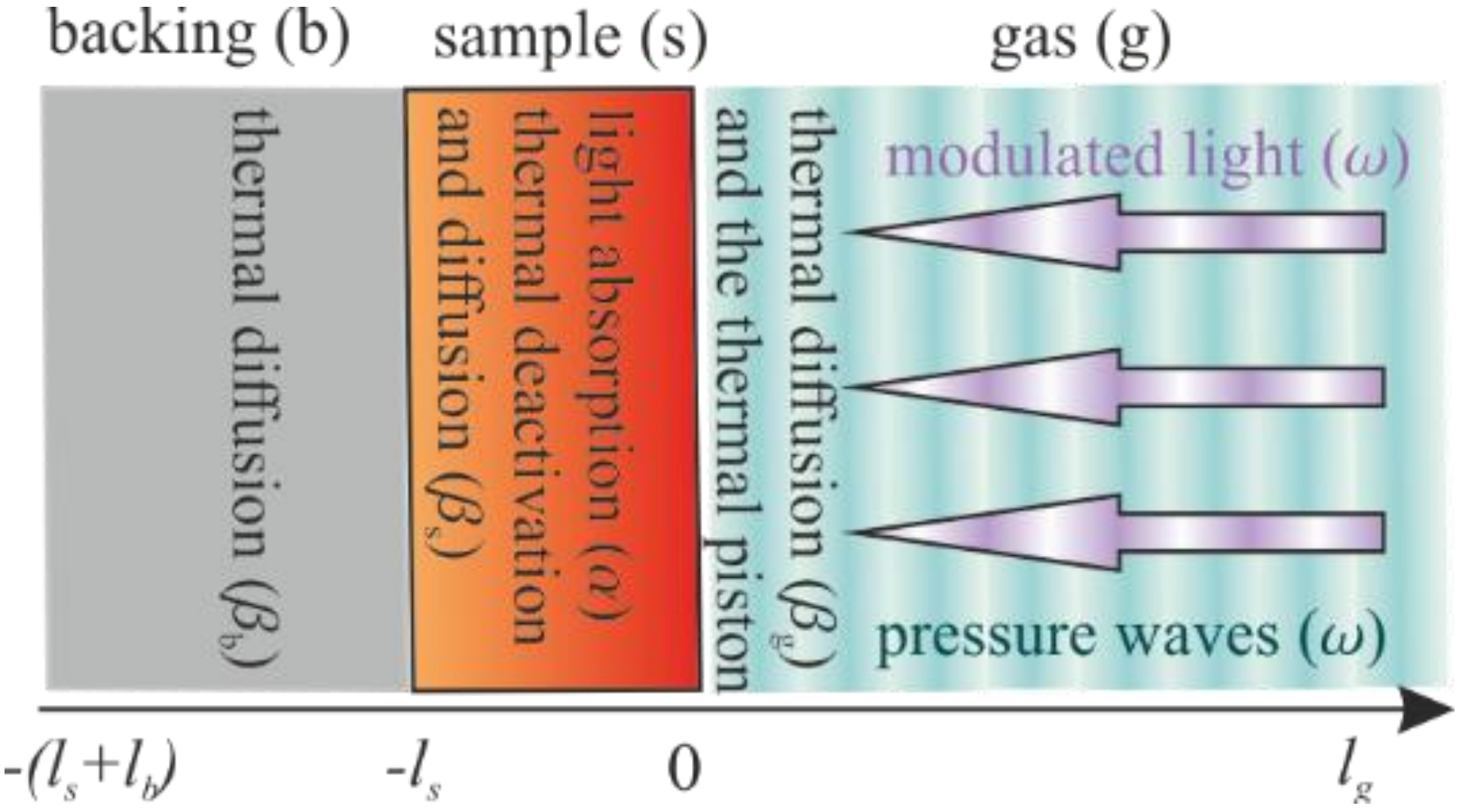

Fig. 2 Cross-section of a simple PA cell showing the position of the sample, backing and gas column, along with the processes leading to the PA signal generation: absorption of the modulated ($\omega$) energy flux by a sample (governed by the Beer-Lambert law), thermal deactivation of a sample and evolution of the perturbed temperature field in the system (governed by the Fourier-Kirchhoff law), generation of the pressure (acoustic) waves ($\omega$) in the gas phase by the action of the thermal piston.

Let us consider sinusoidally modulated monochromatic light of intensity:

$$I = 1/2\, I_0\, (1 + \cos[\omega t]), \qquad (1)$$

where $\omega$ specifies modulation frequency (in angular units, with $\omega = 2\pi f$), incident on a sample. The light absorption by the sample is governed by the Beer-Lambert's law, and leads, via the non-radiative deexcitation, to a certain distribution of heat sources $h$ within the sample:

$$h \equiv h(x,t) = I\alpha \exp[-\alpha x] \qquad \text{for } x \in < -l_s; 0 >, \qquad (2)$$

where $\alpha$ stands for the optical absorption coefficient. It is assumed, that only the sample absorbs the optical energy.

The perturbed temperature field in the sample leads to the heat flow in the system; the process is described by a set of Fourier-Kirchhoff equations:

$$\frac{1}{\beta_s}\frac{\partial \phi}{\partial t} = \frac{\partial^2 \phi}{\partial x^2} + \frac{h}{k_s} \qquad \text{for } -l_s \leq x \leq 0, \qquad (3a)$$

$$\frac{1}{\beta_b}\frac{\partial \phi}{\partial t} = \frac{\partial^2 \phi}{\partial x^2} \qquad \text{for } -(l_s + l_b) \leq x < -l_s, \qquad (3b)$$

$$\frac{1}{\beta_g}\frac{\partial \phi}{\partial t} = \frac{\partial^2 \phi}{\partial x^2} \qquad \text{for } 0 < x \leq l_g, \qquad (3c)$$

where $\phi$ is the temperature (the *physical* temperature is given by: $T = Re[\phi] + ambient$), while $\beta_j$ (*j* = *s*, *b*, *g*) stand for the thermal diffusion coefficients of the sample, backing and gas, and $k_s$ refers to the thermal conductivity of the sample.

The solution of the Eqs 3, which provides the temperature distribution in the system, is generally complicated. For the region of interest - a *thin* gas layer of thickness of $\lambda_g = 2\pi\mu_g$ adjanced to the absorbing sample ($\lambda_g$ is considered as the boundary layer capable of responding thermally to the temperature variations at the sample surface), where $\mu_j = \sqrt{\frac{2\beta_j}{\omega}}$ is the thermal diffusion length, the spatially-averaged temperature variations due to the periodic heating of the layer are described by:

$$\bar{\phi}(t) = \frac{1}{\lambda_g}\int_0^{\lambda_g} \phi_{periodic}\,(x,t)dx \cong \frac{1}{2\sqrt{2}\,\pi}\theta \exp[i(\omega t - \pi/4)], \qquad (4)$$

where $\phi_{periodic}$ denotes the periodic solution of Eqs 3, while $\theta$ is the amplitude of $\phi_{periodic}$ at the sample/gas boundary, and encodes geometrical, optical and thermal properties of the system.

Due to the periodic heating of the boundary layer, the layer expands and contracts periodically; as such, it can be considered as a piston, which acts on the rest of the gas column. Using the ideal gas law, the displacement of the piston can be given as:

$$\delta x = \lambda_g \frac{\overline{\phi}(t)}{T_0} = \frac{\theta \mu_g}{\sqrt{2} T_0} \exp[i(\omega t - \pi/4)]. \quad (5)$$

Assuming that the gas column responds to the piston adiabatically, the incremental pressure is:

$$\delta p = \gamma \frac{p_0}{l_g} \delta x = Q \exp[i(\omega t - \pi/4)], \quad (6)$$

where $\gamma$ is the adiabatic index, while $p_0$ is the ambient pressure, and $Q = \frac{\gamma p_0 \theta \mu_g}{\sqrt{2} l_g T_0}$ stands for the complex envelope of the sinusoidal pressure variations; then, the physical pressure, $\Delta p$, is given by the real part of Eq.6:

$$\Delta p \equiv Re \left\{ Q \exp\left[ i \left( \omega t - \frac{\pi}{4} \right) \right] \right\} = q \cos(\omega t - \psi - \pi/4) \quad (7)$$

where $q$ and $\psi$ specify the magnitude and phase of $Q$. Finally, it is assumed, that the experimentally acquired signal, $S$, is proportional to the physical pressure changes in the chamber. As such, the signal represents all of the sample, backing and gas characteristics encoded in $Q$; thus $S \equiv S(s, b, g, \omega^{-n})$, where $n \epsilon < 1; 1.5 >$ and depends on the geometrical, optical and thermal properties of the sample.

Concluding, according to the RG model, the photoacoustic sample response depends on an interplay between the sample geometrical (thickness $l_s$) and optical properties (absorption coefficient), and the thermal characteristics of the system, provided by the thermal diffusion coefficients of the sample, the backing and the surrounding gas. These characteristics can be further translated into a path-length scale with the relationship for thermal diffusion depths $\mu_j = \sqrt{\frac{2\beta_j}{\omega}} = \sqrt{\frac{\beta_j}{\pi f}}$ ($j$ = $s$, $b$, $g$), and optical absorption length $\mu_\alpha = 1/\alpha$. Using the path-length scale system characteristics, Rosencwaig and Gersho examined special cases of the complex gas pressure variations due to the photothermal sample response. These included optically transparent and opaque solids ($\mu_\alpha > l_s$ and $\mu_\alpha < l_s$, respectively) and thermally thin and thick solids ($\mu_s > l_s$, $\mu_s < l_s$, respectively) with an additional relationship (> or <) between $\mu_s$ and $\mu_\alpha$. Some interesting predictions of the RG theory, especially in relation to the motivation of the paper (see Fig. 1), emerged for thermally thick samples with $\mu_s < \mu_\alpha$. For such samples, due to the thermal wave attenuation (with a factor of $\exp(-x/\mu_s)$), only a certain sample layer, limited by the thermal diffusion depth (related to the material thermal diffusion coefficient) contributes to the PA signal. In other words, only the light absorbed within the first thermal diffusion length ($\mu_s$) contributes to the signal. Eventually, by assuming the correspondence of: $\mu_s = \sqrt{\frac{\beta_s}{\pi f}} \equiv l$, where $l$ denotes the sampling depth, it is possible to scan (or profile) a sample *layer-by-layer* simply by varying the light modulation frequency as long, as the thermal diffusion coefficient of the sample is known.

## 3. Materials and methods

The multibeam approach was validated upon three model samples: (*I*) an optically opaque system – here, a perfect absorber, carbon black membrane (as delivered by MTEC Photoacoustics), (*II*) an optically opaque multilayer system consisting of a copper base plate (CU-M-02-SAMP, American Elements) of thickness $l_s$ = 38 ± 1 μm coated with a commercial red acrylic paint (Baufix, Holz & Bautechnik GMBH, layer thickness of 30 ± 1 μm), and (*III*) a luminescent sample - a ruby ($Al_2O_3:Cr^{3+}$) disc of thickness 2.00 ± 0.01 mm. For all of the measurements, the samples were sealed in a photoacoustic chamber. Prior to the PADP/MBPA measurements, the samples were characterized by PAS at modulation frequency of 120 Hz. The results for samples *II* and *III* (both normalized with respect to the carbon black sample) are shown in Fig. 3a. The PAS characterization allowed to point at particular spectral regions where the samples convert the absorbed light into heat the most efficiently. The light sources for the MBPA were selected to match these characteristics (mainly the absorption bands) of the samples.

The experimental rig for the PAS measurements consisted of a 900 W xenon light source (Newport 66921), grating monochromator with adjacent mechanical chopper (Newport 74125), PAC300 photoacoustic chamber (MTEC Photoacoustics) and SR7265 lock-in amplifier.

Appropriate light sources were selected for the PADP/MBPA studies depending on the spectral characteristics of the samples being studied. These included a pair of LEDs for the validation studies (sections 4.1 and 4.2), and a pair consisting of a laser source and a monochromatic light from a xenon lamp for photosynthesis efficiency studies.

The MBPA setup consisted of two LED modules developed at CMSD UG and controlled by a Siglent 2042X function generator. The modules were equipped with 420 nm and 535 nm Lumileds Luxeons (with peak intensities of 20 and 7 W m$^{-2}$, respectively). The driving voltage range for the light sources was adjusted to maintain LED linear response (light intensity pattern match exactly the pattern provided by the function generator). The scanning beams were routed to PAC 300 in one of two qualitatively equivalent ways: via Y-type multicore fiber or directly via optical cage systems (Fig. 3b); the PA chamber output signal was split into two independent channels with SR7265 unit in each line. A complete block diagram of the PADP/MBPA setup is shown in Fig. 3c.

The pilotage photosynthesis efficiency studies were carried out on the abaxial side of *ficus benjamina L.* samples cut from a fresh leaf collected prior the measurement. The samples were sealed in PAC 300 cells for the measurements. The light sources for the photosynthesis studies were the 658 nm laser source LDCU5 (Power Technology) and the 700 nm light from the xenon source (the same in spectroscopic studies); the photosynthesis-saturating white light (photon flux of 2000 μmol $m^{-2}$ $s^{-1}$) was generated by Dolan-Jenner Fiber-Lite DC-950 illuminator.

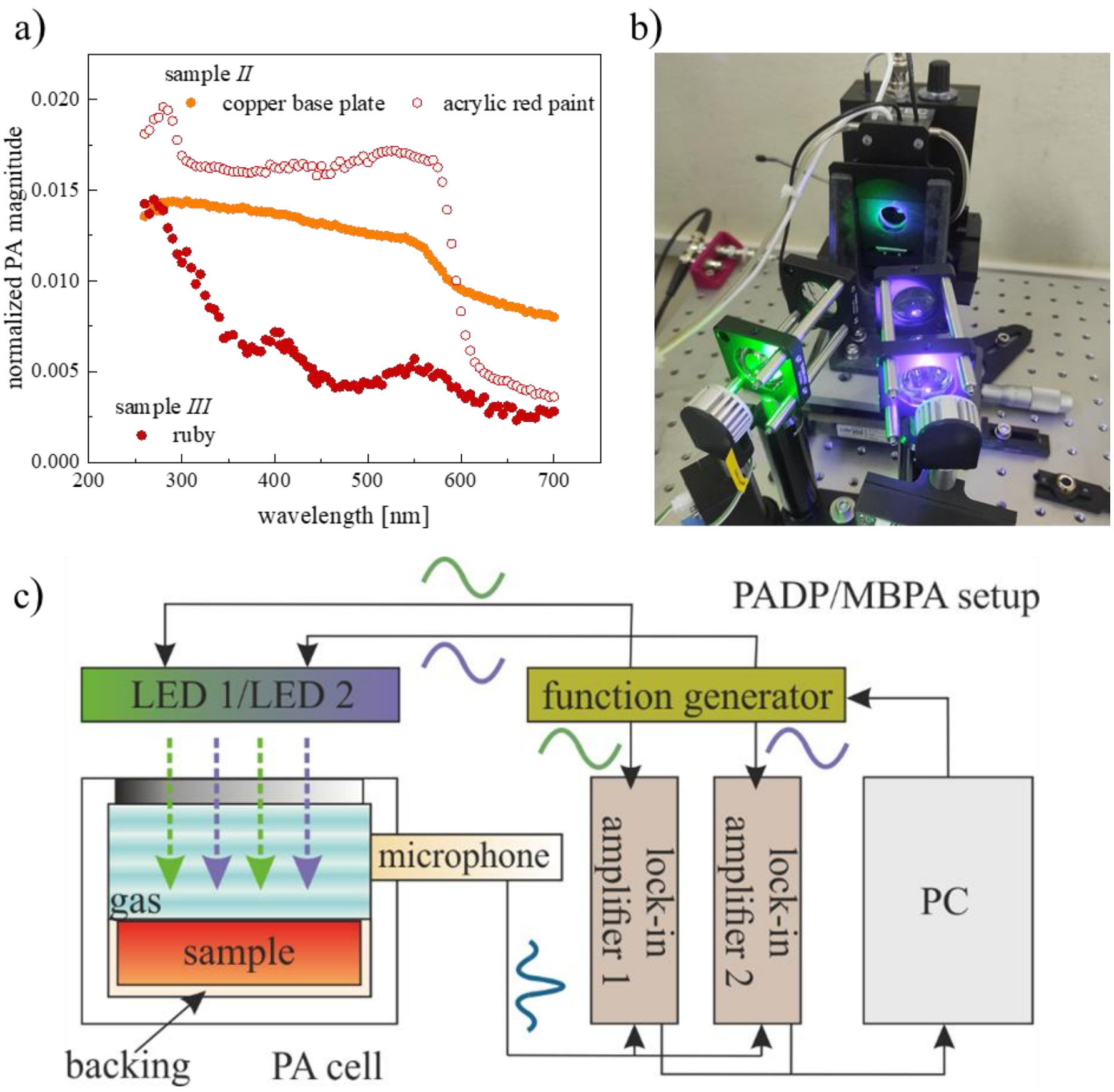

Fig. 3 a) PA spectra of samples *II* and *III*, collected at the modulation frequency of 120 Hz; b) exemplary PADP/MBPA optical setup as used in the validation studies (section 4.1 and 4.2), consisting of two LED sources in a cage system, in front of a sample enclosed in PAC 300 photoacoustic cell (black chamber, in back), c) a simplified block diagram of the PADP/MBPA setup with optical system as in b), with an absorbing sample (illuminated by two light sources controlled by a function generator) enclosed in a PA cell filled with air (pressure waves in the gas phase marked by periodic shades of blue) and equipped with a microphone (front-side detection scheme), with signal split into two lock-in amplifiers controlled with a PC.

Light intensities were measured with the HD 2102.1 photo-radiometer equipped with the LP 471 RAD probe (Delta Ohm). The PADP/MBPA and PAS experiments had the same geometry (same chamber/front side illumination and detection). All the measurements were performed within the modulation frequency range spanning from tens up to 900 Hz. In this frequency range, the microphone response is linear, and the cell do not exhibit any resonances. The raw signals were recorded directly from the PAC by USB-6216 DAQ device (National Instruments).

## 4. Results and discussion

The aim of the studies was to verify the reliability of the *f*-domain MBPA experiments, in the special case of two simultaneous probe beams of different wavelengths. Special attention was paid to the investigation of two crucial issues: I) finding the condition for the modulation frequencies of the two probe beams, that allows for the simultaneous acquisition of two independent PA signals; II) verifying the PA response of a sample to the double-beam excitation character: a single sum of two independent (separately acquired) signals vs a signal composed of single components enriched by synergistic/cross effect contribution. To address these issues, two independent tests were conducted, including raw acoustic signals and comparative analyses of PADPs vs MBPA data.

### 4.1 Raw acoustic signal analyses

Raw signal analysis provides direct access to the overall system responses to incident excitation. By applying the Fast Fourier Transform to the time-domain acoustic response of a sample to a sinusoidally modulated energy flux (as recorded by a DAQ device), it is possible to analyse the overall spectral characteristics of the signal.

By analysing Eqs 1 and 7 it can be concluded, that the fundamental frequency of the acoustic response of a sample to a single beam excitation matches the modulation frequency of the incident light. In fact, the acoustic response reflects periodic temperature variations within the sample being tested - the temperature field exhibits wave-like properties (so called *thermal waves*). It is therefore worth checking whether multi-beam excitation will disrupt the original frequency patterns.

The raw signal analyses were performed for all of the model samples, excited in single and multibeam modes, at various light intensity ratios and multiple arbitrary modulation frequencies below 900 Hz. The signals were recorded for about ~ 5 sec, the spectral resolution (after FFT) was about 0.15 Hz. Exemplary results for the carbon black sample (highest signal-to-noise ratio for all of the samples considered), for arbitrary modulation frequencies of 29 and 39 Hz, are shown in Fig. 4.

Figs 4 a1–b1 demonstrate PA responses of the sample to two separate beams (single excitation at a time) with modulation frequencies of 29 and 39 Hz and magnitude ratio ~ 3/1, and corresponding spectral densities (FFT magnitudes). One can notice a fine agreement with the theoretical predictions of the frequency relation between excitation modulation and PA response. No other spectral features, including higher harmonics or electrical hum signature (~ 50 Hz), are not present. Applying the best-fit of a sharp Lorentz-type function to the data gives narrow FWHMs (*full width at half maximum*, considered here as a measure of frequency dispersion around its *fundamental* value) of ~ 0.2 Hz, close to the spectral resolution of the data.

Figs 4 a2-b2 demonstrate the response of the sample to simultaneous two-beam excitation (same characteristics as in a1-b1: 29 and 39 Hz and magnitude ratio ~ 3/1) in time domain and corresponding spectral densities. As before, the spectral densities remain similar to those analysed for the single beam case: there are no harmonics, nor are there any beat frequencies or other synergistic or cross effects that could potentially affect the signal. There is a slight decrease in the amplitudes and an increase in the FWHMs but, as before, the FWHMs may be influenced by the data resolution. Similar conclusions can be drawn from the analyses of Figs 4 a3-b3, where the two simultaneous beams are of similar magnitude. It should be noted that the issue of small magnitude variations can be further investigated either by analysing longer acoustic response recordings (here the analysis was based on ~5 sec recordings), or by a comparative analysis of PADP vs. MBPA. The latter is presented in the next section.

Similar results were obtained for all of the samples, for different but arbitrary configurations of the beam intensity ratios and modulation frequency shifts $\Delta f = f_1 - f_2$ (in general, the acoustic responses were tested for $\Delta f > 5$ Hz, up to $f_1 = 900$ Hz, which yields $\Delta f / f_1 \approx 0.005$ considered as a slight frequency difference between the sources). From the analysis of the FWHMs, it can be concluded that the minimum modulation frequency separation of the excitation beams, which allows independent sample responses to be obtained, should be greater than 0.4 Hz.

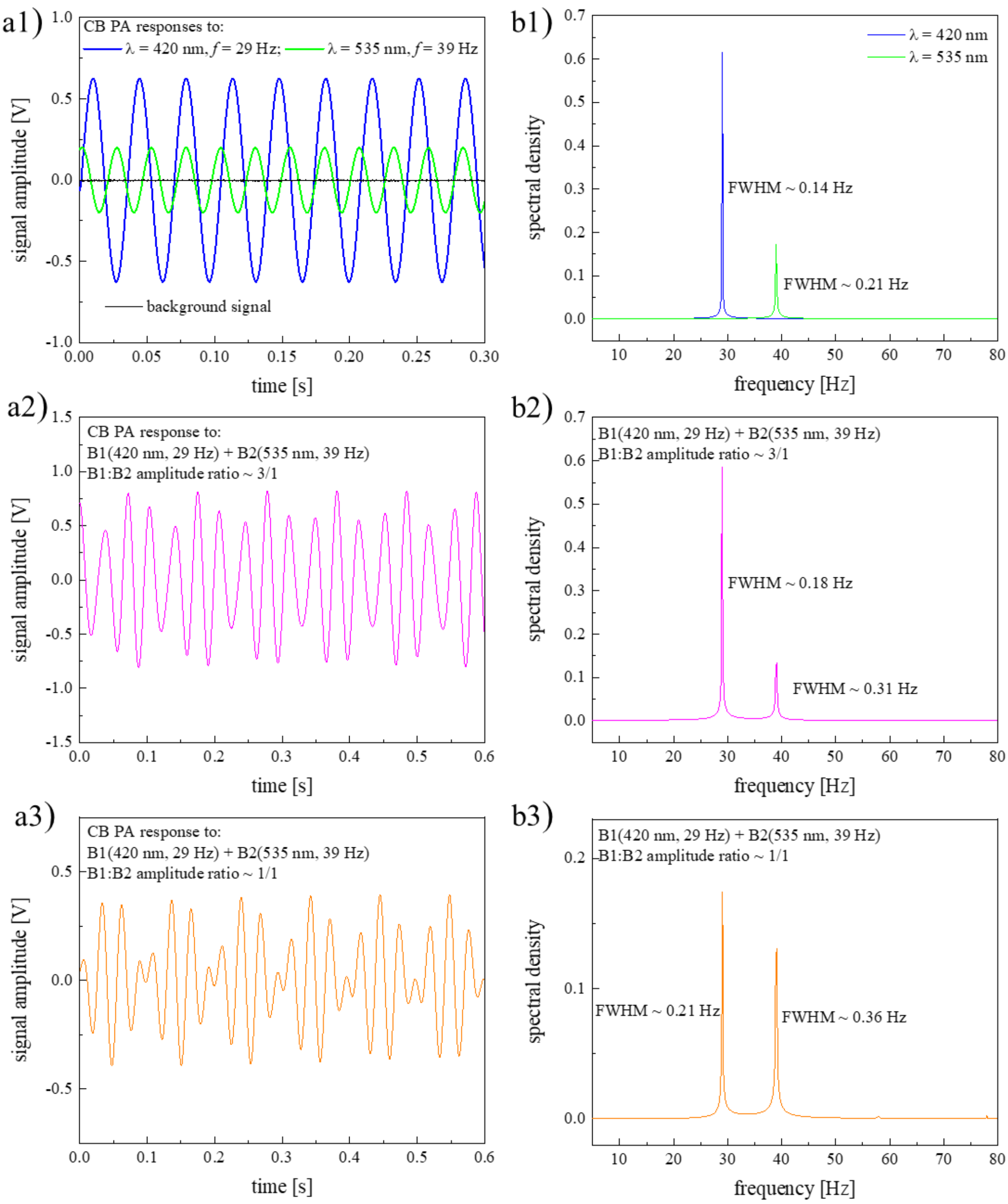

Fig. 4 Exemplary PA responses of carbon black sample to two modulated beams B1 (420 nm, 29 Hz) and B2 (535 nm, 39 Hz) – left column, and corresponding FFTs of the PA responses – right column. Case 1 – PADP – one beam at a time, case 2 – MBPA – two simultaneous beams of amplitude ratio ~3/1, case 3 – MBPA, amplitude ratio ~ 1:1. The modulation frequencies were arbitrary chosen, the frequency ranges were picked to showcase potential beat effects, harmonics etc.

### 4.2 Frequency domain PADP vs MBPA analyses for stationary systems

The same issues, namely experimental conditions for simultaneous detection of independent responses and verification of the presence of potential cross-effects presence, were considered for lock-in photoacoustic measurements.

The first issue was verified in an experiment involving carbon black sample (a perfect absorber). It was found, that two PA responses to simultaneous double beam excitations (420 and 535 nm) characterized by different modulation frequencies $f_1$ and $f_2$ are independent as long, as the modulation frequency shift between the beams ($\Delta f = f_1 - f_2$) exceeds instrumental noise and small frequency oscillations registered by lock-in amplifiers. For the experimental rig used, the magnitude of such oscillations was not greater than ~ 0.2 Hz over the entire range considered (from 15 to 900 Hz), and for $\Delta f < 0.4$ Hz the acquired PA signals appeared to be unstable. These observations are consistent with the findings of the raw acoustic signal analyses. Although the recorded PA responses were generally stable for $\Delta f$ above the sum of the hums from both channels were generally stable, the frequency shift was set to 2 Hz for the following experiments.

To compare the *f*-domain MBPA and independent PADPs, a comparative study was performed for a model CB sample in the $f_1$ frequency range of 16-325 Hz, with the MBPA frequency shift of $\Delta f = 2$ Hz (in case of PADPs, $\Delta f = 0$). In addition, the intensities of the probing beams were adjusted so that the ratio of the recorded PA signals was ~10:1. The results of the experiment are shown in Fig. 5a. It can be noticed, that MBPA responses to 420 and 535 nm excitation follow a similar pattern to the PADP data. A supplementary MBPA experiment was also performed, where the cumulative PA response to two probing beams (with $\Delta f = 0$) was registered. A comparison between the sums of the MBPA signals and PADP signals is shown in Fig. 5a inset; again, a similarity between the profiles can be noticed. Concluding, by incorporating an appropriate frequency shift between two probing beams, it is possible to simultaneously acquire two independent PA responses to distinct excitation wavelengths (an equivalence between MBPA and single PADP signals); one does not observe any influence of the beams intensity (at least up to the ratio of 10:1) or modulation frequency shifts ($\Delta f / f_1$, at least down to value of ~ 0.0025) on MBPA signals. Moreover, under the experimental conditions as described, the PA response to the double beam excitation appears to be given by a sum of individual contributions (no non-linear or cross-effects during the PA signal generation/detection were observed).

Similar conclusions regarding the PA signal magnitudes can be drawn from the results for the other samples: a multilayer system - acrylic red paint-coated copper plate (Fig. 5b, $f_1 \epsilon (16; 80)$ Hz; for clarity, only the PADP/MBPA response to 420 nm beams is shown) and a luminescent sample - $Al_2O_3:Cr^{3+}$ (Fig. 5b inset, $f_1 \epsilon (16; 900)$ Hz, signal in μV), where the beams were adjusted to give a signal ratio of 1:2 and 1:1, respectively. In all cases considered, the signal phase behaviour for PADP and MBPA was also similar (observed deviations are considered to be within ±1° experimental error).

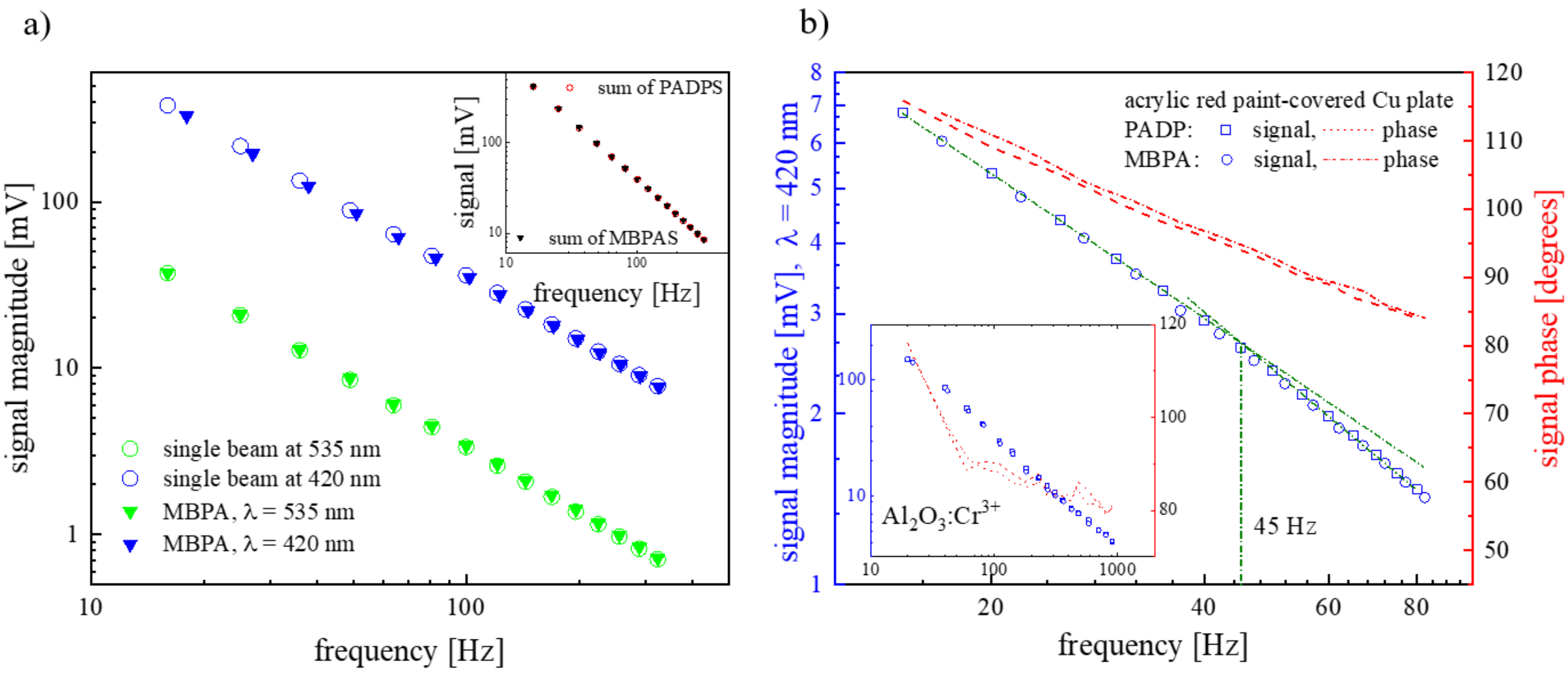

Fig. 5 Comparison of PADP and MBPA results for: a) carbon black sample (inset – cumulative MBPA and PADP responses for $\Delta f = 0$), b) a multilayer sample II (main figure, along with a breakthrough frequency of 45 Hz) and a luminescent sample III (inset).

Photothermal techniques are considered the *methods-of-choice* for the characterization of thermal properties of materials. In case of layered samples, a very basic approach for the layer thermal diffusivity (*β)* quantification involves the determination of a characteristic cut-off frequency $f_c$ (Fig. 5b – intersection of the linear fits to the low and high frequency data), related to the measureable layer thickness $l$ (assumed to be equal to the thermal diffusion length, $\mu_s$) via: $\beta = l^2 \Pi f_c$. In the case of sample II (a copper plate covered with acrylic paint), both PADP and MBPA data revealed $f_c$ to be ~ 45 Hz (corresponding to the paint layer of thickness of $l$ = 30 ± 1 μm). Thus, a rough approximation (the $f_c$-based methodology applied is considered as a basic one; a more advanced procedure can found in [16]) of the thermal diffusivity of the acrylic paint (the outer layer) was found to be (1.30 ± 0.04)·$10^{-3}$ $cm^2$ $s^{-1}$, which stands in agreement with the literature data [17]. In general, the two methods (PADP and MBPA) seem to give similar results for stationary systems.

Last but not least: the relationship between the sampling depth and frequency is given by $l \propto f^{-1/2}$, which indicates monotonically increasing frequency steps to maintain a constant sampling depth interval. For higher modulation frequencies, the frequency shift is of minor effect when analysing systems, f.e. for water system example described in the introduction, for modulation frequencies of ~10 Hz (sampling depth of 67 μm) the 0.4 Hz frequency shift leads to a difference in the sampled layers thickness of 1 μm (1.5% of the total layer thickness), and for ~1 kHz (6.7 μm) the difference drops to $10^{-3}$ μm (0.015% of the layer). As such, the magnitude of the frequency shift of 0.4 Hz appears to be of minor importance for applied science measurements.

4.3 Remarks on the MBPA signal generation

Let us highlight some of the findings on MBPA signal generation in relation to the RG model outlined earlier. In general, we consider the signal generation in the presence of two beams with modulation frequencies of $\omega_1$ and $\omega_2$ and wavelengths $\lambda_1$ and $\lambda_2$.

First, the excitation is due to a sum of beams characterized by two modulation frequencies $\omega_1 \neq \omega_2$, and the system is characterised by absorption coefficients such that: $l_s > 1/\alpha_1 \sim 1/\alpha_2$. Then, Eq. 1 can be replaced by:

$$I_T = 1/2\, I_{0,1}\, (1 + \cos[\omega_1 t]) + 1/2\, I_{0,2}\, (1 + \cos[\omega_2 t]), \qquad (8)$$

where $I_T$ refers to the total light intensity. The sample temperature field perturbation due to the absorption of the beams is given by the heat equation in the form:

$$\frac{1}{\beta_s}\frac{\partial \phi}{\partial t} = \frac{\partial^2 \phi}{\partial x^2} + \frac{\alpha_1 \exp[-\alpha_1 x]}{k_s} I_T. \qquad (9)$$

Due to linearity of the RG model, one expects a complex solution with two Fourier components of $\exp[i\omega_1 t]$ and $\exp[i\omega_2 t]$, and the system response as a function of $\cos[\omega_1 t + \phi_1]$ and $\cos[\omega_2 t + \phi_2]$, where $\phi$ denotes the signal phase (see Eq. 7). A similar case was studied in Fig. 3.3 (same signals intensities, different modulation frequencies). The MBPA appeared to be equal to the sum of PADPs, no additional effects were observed.

Second, if the $\omega_1 = \omega_2$ and $\alpha_1 \neq \alpha_2$, then the heat source distribution is given by:

$$h_T \equiv h_1(x,t) + h_2(x,t) = I_1(\alpha_1 \exp[-\alpha_1 x] + \alpha_2 \exp[-\alpha_2 x]), \qquad (10)$$

and thus the fundamental solution of the RG theory display two sets of components with exponential-type spatial dependences. A somewhat similar case was studied in Fig.4a (same frequencies and absorption coefficient, but different light intensities). The MBPA signal appeared to be equal the sum of PADPs.

In addition, by taking advantage of the function generator's functionality of phase shifting, two beams ($\omega_1 = \omega_2$, $I_{0,1}\, \alpha_1 = I_{0,2}\, \alpha_2$) were generated in counterphase (data not shown). This led to the signal drop down to zero, which implies a simple destructive interference pattern of two waves shifted by $\pi$.

Other experimental cases considered here involved *mixed* conditions, i.e. where $\omega_1 \neq \omega_2$, $\alpha_1 \neq \alpha_2$ and $I_{0,1}\,\alpha_1 \neq I_{0,2}\,\alpha_2$, which could potentially give rise to some cross effects. By analysing the results obtained one can draw a conclusion, that the information encoded in MBPA signal is equivalent to the sum of the PADP responses. It appears that individual contributions to the overall MBPA signal can be analysed using linear PA signal generation models developed for single beam excitation, at least for absorbing systems under weak irradiation conditions and relatively low frequency range, as studied here. A complete theoretical framework for MBPA signal generation would be beneficial to guide potentially interesting cross effects not recognised in these preliminary studies.

### 4.4 Special case – photosynthetic energy storage analyses

As mentioned in the introduction, the applicability of photothermal analysis extends far beyond the study of inanimate matter. An example is the measurement of photosynthetic activity, which was established shortly after the development of the Rosencwaig-Gersho theory [18,19]. Recently, photoacoustic sensors have been shown to be suitable for a broad range of water biomonitoring, including microplastic and chemical presence detection, together with photosynthetic energy storage (PS) efficiency analyses as an additional pollution biomarker [5].

The PS efficiency measurements exceed the theoretical PA background provided by the RG model. First, the overall signal is composed of photothermal and photobaric contributions; the latter one is related to the oxygen evolution of a photosynthetic-active sample. The photobaric contribution is usually time-shifted with respect to thermal response; due to a phase lag between the contributions, the overall PA response can be attenuated or enhanced relative to its photothermal component. Secondly, the photothermal contribution can be reduced by effects related to the photochemical loss or photosynthetic energy storage – a measure of photosynthetic activity.

To quantify the PS efficiency, two modifications to the standard PADP experiment (as in section 4.2) must be made. First, the experiment has to be performed in the frequency range exceeding the reciprocal of the oxygen evolution characteristic time; then, the photobaric component of the signal can be neglected. Secondly, by using a photosynthesis-saturating unmodulated background light, it is possible to reduce the effect of the energy storage effect on the photothermal response. Finally, by performing an experiment within a suitable modulation frequency range with two light sources (modulated and non-modulated), the energy storage parameter can be derived from $ES = 1 - PA_{DARK}/PA_{SAT}$, where $PA_{DARK}$ and $PA_{SAT}$ are the signals registered in the absence and presence of the background light, respectively [5].

Exemplary data collected for the abaxial side of a *Ficus benjamina L.* leaf, highlighting the aforementioned effects but also a rich light-harvesting pigment composition of the sample, including chlorophylls and carotenoids [20], are shown in Fig. 6. For the experiments, 3 round samples of ~ 1 cm diameter were cut from a *ficus benjamina L.* leaf collected on the same day. The first sample was studied by PAS (UV/VIS range, three modulation frequencies, measurement time (sample sealed in chamber) ~ 30 min), while the other two by PADP (two scans in the frequency range of 30 - 210 Hz, single probe beam, measurement time ~ 10 min) and MBPA (double-shot and double-beam measurements in the presence and absence of the background light, measurement time ~ 1 min). For the first sample (Fig. 6a), the PA signal decreases with increasing frequency due to the frequency-limited sampling depth (as in stationary systems) and oxygen evolution effects, while the effect of saturating/non-modulated light presence depends on the scanning light frequency. However, above the oxygen evolution cut-off frequency, the relation between the signals $(PA_{DARK}/PA_{SAT})$ is expected to approach a constant value. Here, the 680 nm band, corresponding to the maximum absorption of chlorophyll a (Fig. 6a), was selected to investigate the $1 - PA_{DARK}/PA_{SAT}$ vs frequency behaviour. The results are shown in Fig. 6b.

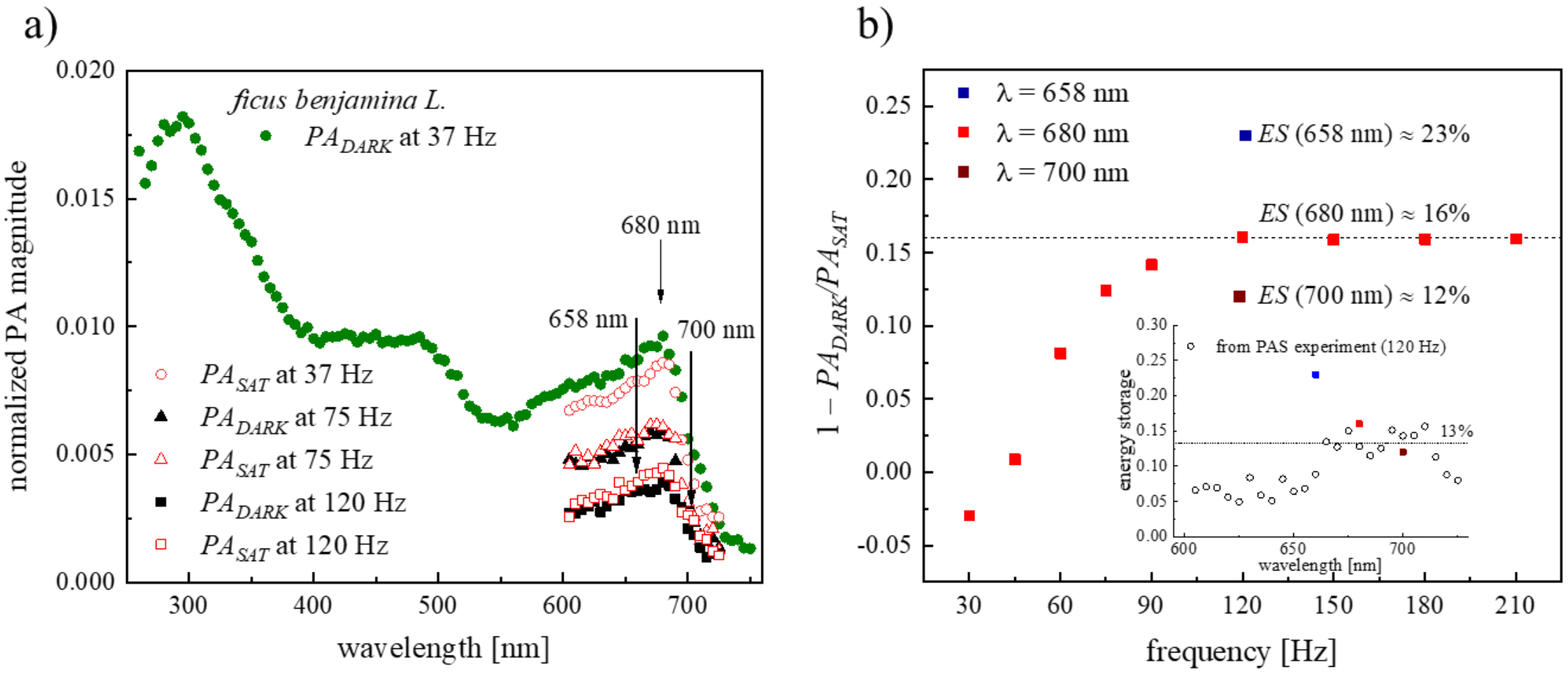

Fig. 6 a) photoacoustic *dark* spectrum of abaxial side of a *Ficus benjamina L.* leaf recorded in 260 – 750 nm range (green full circles) and the *saturated* PAS spectrum in 600 – 720 nm range (red circles) recorded at 37 Hz, along with the *dark* and *saturated* PAS spectra in the chlorophyll red band region at higher frequencies; b) the $1 - PA_{DARK}/PA_{SAT}$ ratio vs modulation frequency recorded for a single 680 nm excitation (PADP, red squares) along with MBPA data – the simultaneous energy storage efficiency quantification for distinct absorption sites (658 and 700 nm at 121 and 119 Hz, respectively). The inset illustrates comparison of the ES (modulation frequency ~120 Hz) quantified upon PAS (open circles) and PADP/MBPA (squares) data.

It is clear, that the asymptotic value of the ratio is reached at ~ 120 Hz; therefore, the *energy storage* efficiency for the 680 nm probe beam can be estimated to be 16%. On the other hand, it is known that the broad chlorophyll absorption band in the red region can be attributed to the photosystems I and II (PSI and PSII) presence. The photosystems, which are protein complexes located in the thylakoid membranes of chloroplasts, are characterized by a distinct main absorption maxima, typically of 700 nm (PSI), and 680 nm (PSII) [21]. By taking advantage of the sample red region spectra, the cut-off frequency of ~ 120 Hz, and the multibeam setup (here: two scanning beams and saturating light) it is possible to perform a pilotage studies on the energy storage efficiency quantification in the regions corresponding to PSI and PSII bands simultaneously. The experiment was performed on a fresh leaf sample, with two simultaneous probe beams of 658 nm (considered here as a region mainly influenced by PSII) and 700 nm (supposed to be dominated by PSI) [22,23] in the presence and absence of the saturating light (in fact, it was a fast, double shot-experiment); the corresponding modulation frequencies of the probing beams were 121 and 119 Hz, fulfilling the modulation shift criterion. The energy storage values obtained, *ES(658 nm)* = 23% and *ES(700 nm)* = 12%, differ significantly from each other and the reference value of *ES(680 nm)* = 16% reported above for a single probe beam experiment, and found during the PAS experiment (~ 13% on average in the range of 660 - 700 nm), all summarized in the inset of Fig. 6b. The MBPA results, at least qualitatively, stay in agreement with literature data for sugar maple leaves ($ES(PSI)/ES(PSII) \approx 1/3$) in, so called, state 1, i.e. when PSI receives too much excitation energy from the background light (the state transitions are acclimation mechanism that rebalances excitation pressure on the photosystems) [24,25].

Finally, the importance of multi-beam experiments for the quantification of PSI/PSII energy storage efficiency should be emphasised. Typically, wavelength dependent ES quantification (allowing PSI/PSII differentiation) involves two sample scans over a given spectral region (and a constant modulation frequency above the oxygen evolution cut-off frequency). Due to the PA signal stabilization, the acquisition of both spectra ($PA_{DARK}(\lambda)$ and $PA_{SAT}(\lambda)$) is rather time consuming; for instance, during our previous studies on the *ES* quantification of marine samples: for 400 – 700 nm region with 5 nm resolution (60 points), the acquisition time-constant of $\tau$ = 2 seconds and the relaxation time of $3\tau$, single scan took 8 minutes [26]. In case of biological systems, the PA measurement time is crucial for two reasons. Firstly, most biological systems are mainly composed of water, which evaporates during the measurements, and the rate of evaporation is increased when the sample is exposed to strong light (here, the white background light). This effect affects the PA signal in two different ways: through changes in the characteristics of the sample (reduced volume/sample thickness and consequent changes in the thermal and optical parameters, which actually disturbs the sample) and through changes in the thermal parameters of the surrounding gas (and possible water condensation in the photoacoustic cell). The latter can be reduced by sufficiently long sample stabilisation under extreme experimental conditions (i.e. in the presence of strong background light affecting the sample properties). The second reason is that biological samples tend to deteriorate away from their natural environment. Therefore, proper quantification of biological processes requires the measurement to be performed shortly after sample extraction/collection. These two reasons point to a fast but reliable identification and quantification of the processes of interest. The use of multibeam technology reduces the measurement time (external sample stressor) in proportion to the number of scanning beams.

## 5. Conclusions and perspectives

The research focuses on experimental tests to validate PA measurements in the presence of two simultaneous frequency-modulated probe beams. The crucial findings indicate, that the PA responses to two modulated beams (of different wavelength) are independent as long, as the modulation frequency shift between the modulation frequencies of the beams exceeds instrumental instability, and the cumulative signal is given by a simple sum of the two PA contributions (no synergistic or cross effects detected). Other possible factors influencing a single PA response in the two-beams configuration (beam intensity ratio, relative modulation frequency shift) were not recognized. The findings appear to be valid at least for absorbing systems ($l > 1/\alpha$) under weak irradiation conditions (LEDs as light sources) and rather low modulation frequencies (microphone range).

It is understood, that the prospects for the MBPA modality are focused on dynamic systems, where simultaneous PA detection/observation of different species/pigment/sites is of special interest, rather than the characterization of stationary systems, where a single beam PA is sufficient. From the author's point of view, MBPA can be applied in areas where a single beam PA is already considered a unique experimental tool. These include:

- transdermal delivery of drugs/membrane transport; as recently shown, PAS and PADP, unlike the majority of methods of choice in this field, allow the identification of the physical processes underlying mass permeation [14]. Until now, only the detection of a single permeant type (characterized by a specific absorption signature) was possible. The application of MBPA can extend the PA potential, and allow for the characterization of multiple permeating species simultaneously. In particular, for the first time quantification of photophysical and photochemical processes, like drug degradation or binding [2], crucial in the field of pharmacy, will be directly accessible;

- photosynthesis (PS) is considered to be one of the most important physical processes. The PA-based quantification of the process efficiency involves PADP studies in the chlorophyll red band region. In fact, the band consists of overlapping bands related to two distinct photosystems. As such, the MBPA may allow PS efficiency quantification to be extended to real-time measurements with additional differentiation between the contributions of the reaction centres; alternatively, an overall PS efficiency could serve as a multi-parameter air/water quality indicator when tested against the presence of other pollutants. The idea for the PA-based water quality sensors, with separate modes for the microplastics detection and PS efficiency quantification, has recently been developed and tested for model systems [5].

**Acknowledgements**

The author thanks M. Grzegorczyk of CMSD UG for inspiring discussions.